\documentclass[sn-mathphys,Numbered]{sn-jnl}

\usepackage{graphicx}
\usepackage{xspace}
\usepackage{multirow}
\usepackage{amsmath,amssymb,amsfonts}
\usepackage{amsthm}
\usepackage{mathrsfs}
\usepackage[title]{appendix}
\usepackage{xcolor}
\usepackage{textcomp}
\usepackage{manyfoot}
\usepackage{booktabs}
\usepackage{algorithm}
\usepackage{algorithmicx}
\usepackage{algpseudocode}
\usepackage{listings}
\usepackage{multirow}
\usepackage{booktabs}
\usepackage{epstopdf}
\usepackage{hyperref}
\usepackage{natbib}

\begin{document}

\title[Article Title]{Identification of Hemorrhage and Infarct Lesions on Brain CT Images using Deep Learning}

\author*[1]{\fnm{Arunkumar} \sur{Govindarajan}}\email{arunkumar@aarthiscans.com}

\author[2]{\fnm{Arjun} \sur{Agarwal}}\email{arjun.agarwal@qure.ai}

\author*[2]{\fnm{Subhankar} \sur{Chattoraj}}\email{subhankar.chattoraj@qure.ai}
\author[2]{\fnm{Dennis} \sur{Robert}}\email{dennis.robert@qure.ai}
\author[2]{\fnm{Satish} \sur{Golla}}\email{satish.golla@qure.ai}
\author[2]{\fnm{Ujjwal} \sur{Upadhyay}}\email{ujjwal.upadhyay@qure.ai}
\author[2]{\fnm{Swetha} \sur{Tanamala}}\email{swetha.tanamala@qure.ai}
\author[1]{\fnm{Aarthi} \sur{Govindarajan}}\email{aarthi@aarthiscans.com}
\affil*[1]{\orgname{Aarthi Scans \& Labs}, \city{Chennai}, \state{Tamil Nadu}, \country{India}}
\affil[2]{\orgname{Qure.ai}, \orgaddress{\city{Mumbai}, \state{Maharashtra}, \country{India}}}


\abstract{Head Non-contrast computed tomography (NCCT) scan remain the preferred primary imaging modality due to their widespread availability and speed. However, the current standard for manual annotations of abnormal brain tissue on head-NCCT scans involves significant disadvantages like \textit{lack of cutoff standardization} and \textit{degeneration identification}. The recent advancement of deep learning-based computer-aided diagnostic (CAD) models in the multidisciplinary domain has created vast opportunities in neurological medical imaging. Significant literature has been published earlier in the automated identification of brain tissue on different imaging modalities. However, determining Intracranial hemorrhage (ICH) and infarct can be challenging due to image texture, volume size, and scan quality variability. This retrospective validation study evaluated a DL-based algorithm identifying Intracranial hemorrhage (ICH) and infarct from head-NCCT scans. The head-NCCT scans dataset was collected consecutively from multiple diagnostic imaging centers across India. The study exhibits the potential and limitations of such DL-based software for introduction in routine workflow in extensive healthcare facilities.}




\maketitle

\section{Introduction}\label{sec1}
In Cognitive Neuroscience, Neuropsychological investigation of stroke patients is widely utilized in advancing our knowledge of brain functions. The considerable insight into the relation of the brain function to its anatomy has been determined via correlation analysis between physical brain damage and impaired behavior \cite{Intro_1}\cite{Intro_2}\cite{Intro_3}. The stroke topology can be broadly classified into two types: $1)$ \textit{Intracranial hemorrhage (ICH)}, the rupture blood vessel within the brain which causes bleeding. The common factors related to the cause of ICH are advanced age, heavy alcohol usage, and high blood pressure (hypertension) \cite{Intro_5}. As per some recent studies, although ICH accounts for 10–15\% of all stroke-related deaths, over the last thirty years, the mortality and morbidity have not changed, particularly in developing countries \cite{Intro_4}.  $2)$ \textit{Ischemic stroke or infarct}, is interruption of blood flow due to blood clot. Infarct is generally caused by the 
buildup of plaques (atherosclerosis) over time in the arteries. Globally, over 13.7 million individuals have a stroke each year, of which approximately 70\%, i.e., 9.5 million, are infarct \cite{Intro_6}. Presently, mapping of the stroke lesion is regularly done using Computed tomography (CT) and magnetic resonance imaging (MRI). The MR (T1- weighted and T2- weighted) anatomical images are acquired as a part of routine practice for stroke patients. In stroke suspected patients with negative CT scans, MRI can also be performed. After the first few hours of onset, the ischemic stroke can be identified using the MRI. Additionally, the differentiation of irreparably damaged brain tissue and the tissue at risk due to infraction can be diagnosed using the MRI. However, CT is the preferred imaging modality over MRI in acute stroke care units and clinical trials due to the reduced exclusion criteria compared to MRI, affordability, speed, and accessibility \cite{Intro_7}. In CT, hemorrhage is percieved as the bright region (hyper-dense) exhibiting sharp contrast and infarct as dark region (hypo-dense) depending on the time progressed after the onset.

The manual annotations of abnormal brain tissue by trained neuroradiologists is currently the present standard method for lesion identification \cite{Intro_8}. However, the manual annotations of abnormal brain tissue have many disadvantages \cite{Intro_9}. $1)$ \textit{Lack of cutoff standardization:}, There is no standard protocol for explicit cutoff, particularly around the ventricles, to differentiate lesioned and non-lesioned tissues; as a result, this approach produces large variability and lacks reproducibility across operators. $2)$ \textit{Degeneration identification:} The stroke-induced degeneration occurring in chronic stroke patients outside the lesion is not captured in the standard manual annotations process, even though a significant clinical impact on patients is caused due to the stroke-induced degeneration. The recent advancement of deep learning based computer aided diagnostic (CAD) models in medical imaging and signal processing can significantly assist in overcoming the existing challenges \cite{Intro_10}\cite{Intro_11}\cite{Intro_12}\cite{Intro_13}\cite{Intro_14}. In addition, the manual editing combined with an automated detection solution of hypo- or hyper-dense regions that remains under operator supervision and can assist in overcoming the present challenges \cite{Intro_15}. More recently, a study using large CT datasets to remove the inter-subject variability in brain lesion characterization using an automated approach was proposed \cite{Intro_16}. Several state-of-the-art algorithms have been proposed for lesion segmentation in MR images over the past few years, but very few have been developed to address stroke lesions on CT scans. Most of the earlier work published to validate automated solutions was directed toward identifying ICH. As the ICH appears bright in CT scans, developing an automated solution based on supervised or unsupervised learning algorithm or extracting morphological features from labeled images to differentiate between true lesioned and non-lesioned tissues is less challenging  \cite{Intro_17} \cite{Intro_18}.
Infarct identification, on the other hand, is a less popular problem statement among researchers compared to ICH detection due to its challenging nature. To address this issue very recently, a rule-based approach based on seeded region-growing algorithms was proposed via extracting hand-crafted features such as relative position for an axis of symmetry, texture, and brightness \cite{Intro_19}. However, the primary disadvantage of this study is that the seeded region-growing algorithms may not be able to define the boundaries of the stroke region distinctively. 

In this study, we have evaluated an Artificial Intelligence (AI) based automated CAD algorithm based on deep learning, capable of identifying ICH and infarct on Head-Non-contrast Computed Tomography (Head-NCCT) scan. The solution has been earlier validated on detecting ICH on Head-NCCT scan images \cite{Intro_14}. The Institutional Review Board (IRB) has approved the proposed retrospective study. We demonstrated the effectiveness and validity of the automated CAD solution in detecting ICH infarct and quantifying infarct on Head-NCCT scan. Our proposed validation will provide a rapid and efficient tool for both research and clinical application. It will assist in the broader adaptation of automated CAD solutions at extensive clinical facilities.
\section{Material and Methods}\label{methods}
The study was a HIPAA-compliant retrospective study with Institutional Review Approval (IRB) from Royal Pune Independent Ethics Committee (RPIEC) (IRB No. RPIEC240123). Informed consent was obtained from all participants. All methods were carried out in accordance with relevant guidelines and regulations.

The primary objective was to evaluate the commercially available deep learning-based algorithm qER (Qure.ai Technologies, Mumbai, India) in terms of Area Under the Receiver Operating Characteristics Curve (AUC) in triaging Head-NCCT scan in detection and quantification of infarcts. It was estimated that a minimum sample of 418 Head-NCCT scans (167 Head-NCCT scans image with radiologist-confirmed infarcts, 251 Head-NCCT scans images without infarcts, 2:3 ratio) would provide a minimum of 80\% power to estimate an anticipated AUC of 80\% with 7\% precision assuming a Type I error rate of 5\% \cite{stats_1}\cite{stats_2}. The Head-NCCT scans, and their signed-off original radiological report performed from 01-September-2021 to 31-August-2022 were acquired from diagnostic imaging centers across India. A total of 1878 Head-NCCT scan were collected. The original radiological report of these scans was subjected to a manual review by a clinical data abstractor to classify the scans into infarct, and non-infarct reported scans based on the original radiological report. A stratified random sample of 500 Head-NCCT scans stratified by the presence and absence of infarct (based on the original radiological reports) were then selected for independent ground truthing by a radiologist with more than fourteen years of experience. The inclusion criteria were Head-NCCT scans with soft reconstruction kernel covering the complete brain, slice thickness $\leq 6mm$. The exclusion criteria were Head-NCCT scans with obvious postoperative defects or from patients who had previously undergone brain surgery, Head-NCCT scans with artifacts such as burr holes, shunts or clips, Head-NCCT scans containing metal artifacts, excessive motion artifacts, Head-NCCT scans containing missing and improperly ordered slices. The ground truther radiologist had access to the original head NCCT scan image but was blinded to the original radiology report. The ground truther reviewed all the Head-NCCT scans and provided segmentation boundaries for infarcts and intracranial hemorrhages. The ground truther radiologist also provided a binary response for the presence or absence of cranial fracture, midline shift, and mass effect. The ground truth output was the reference standard for all downstream statistical analyses, not the original radiological report. 

The sensitivity and specificity were estimated based on a default device threshold (available from the manufacturer based on internal testing), and the optimum threshold was based on Youden's index. The 95\% confidence intervals for sensitivity and specificity are reported based on exact method \cite{Stats_3}. AUC and 95\% confidence interval (CI) was estimated based on the empirical method and De Long methodology, respectively \cite{Stats_4}. The segmentation provided by the ground truther radiologist was utilized for the quantification analysis of the error in the predicted infarct volume by the DL-based algorithm. Absolute errors in infarct volume estimation in milliliter (mL), and summary statistics of absolute errors were reported. The statistical analyses were performed using RStudio (RStudio version 2022.07.1, R version 4.2.1) and Python version 3.9.7.
\begin{table}[!htp]
\centering
\caption{Distribution of Head-NCCT Scan Images used for the Analysis Stratified by Findings as Determined by Ground Truth.}
\label{dataset}
\begin{tabular}{@{}c|l|c|c|l|c@{}}
\midrule
\multicolumn{3}{ c |}{\textbf{Infarct (\%)}} & \multicolumn{3}{ c }{\textbf{Non-Infarct (\%)}}\\ 
\midrule
N (\%) & Other subgroups & n \# & N (\%) & Other subgroups & n \#\\
\midrule
&No other target abnormality* &170 &&No other target abnormality* &212\\
\cmidrule{2-3}
\cmidrule{5-6}
\textbf{187} &ICH &7 &\textbf{241}  &ICH &14\\
\cmidrule{2-3}
\cmidrule{5-6}
(\textbf{43.7\%})&Cranial Fracture &5 &  (\textbf{56.3\%}) &Cranial Fracture &18\\
\cmidrule{2-3}
\cmidrule{5-6}
&Midline Shift &2 &  &Midline Shift &2\\
\cmidrule{2-3}
\cmidrule{5-6}
&Mass Effect&7 &  &Mass Effect &8\\
\midrule
\end{tabular}
\begin{tablenotes}\footnotesize
\item[*] Target abnormalities were custom defined and consisted of infarct, intracranial hemorrhage, cranial fracture, midline shift and mass effect.
\item[\#] Denotes the absolute number of scans in the subgroups, and the numbers will not add up to N because one scan may contain multiple target abnormalities.
\end{tablenotes}
\vspace{-5mm}
\end{table}

\section{Experimental Results}\label{sec2}

\subsection{Identification of ICH and Infarct}
The ground truthing was completed for 428, while 22 Head-NCCT scan were excluded due to the inclusion and exclusion criteria mentioned in section \ref{methods}. A total of 187 Head-NCCT scan confirmed (based on ground truth) the presence, while 241 Head-NCCT scan confirmed the absence of any infarcts. This distribution of scans with and without infarcts met the minimum sample size requirements described earlier in \ref{methods}. In addition, 21 scans with intracranial hemorrhages (ICH) and 23 scans with cranial fractures were present in the sample. A total of 212 (49.5\%) of the 428 Head-NCCT scans did not contain any infarcts, intracranial hemorrhages, cranial fracture, midline shift, or mass effect. The distribution of the Head-NCCT scans is shown in Table. \ref{dataset}.

\begin{figure}[!htp]
\centering
\includegraphics[width=\textwidth]{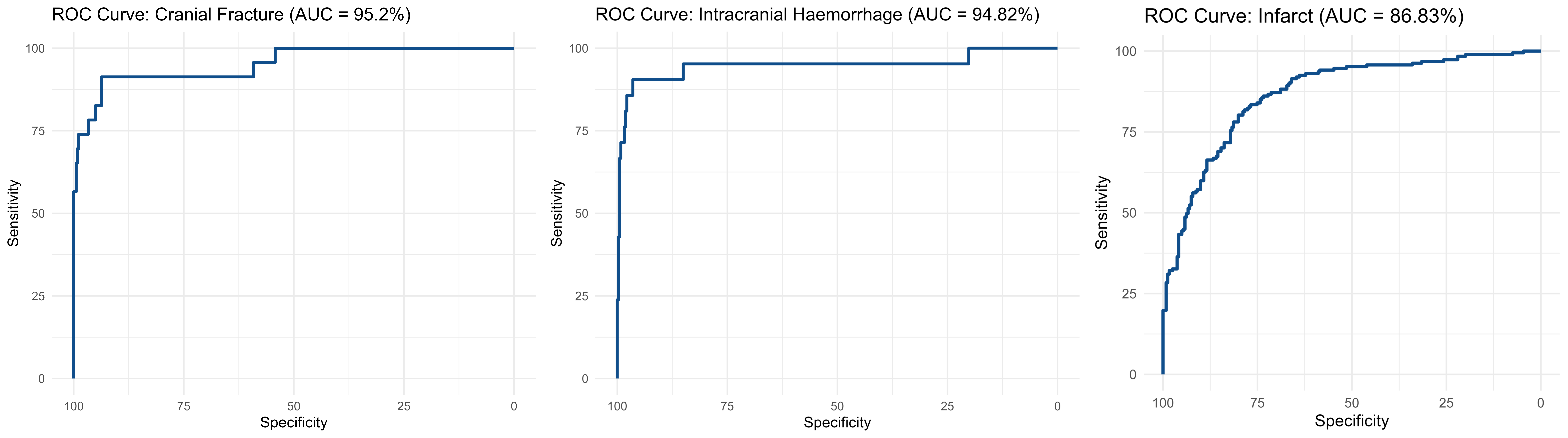}
\caption{AUC-ROC plot: (i) Cranial Fracture (ii) ICH and (iii) Infarct.}
\label{auc_roc_plot}
\end{figure}

It can be observed from Table. \ref{results_binary} that the DL-based algorithm achieved an AUC of 86.8\% (95\% CI: 83.4 - 90.2) in detecting scans with the presence of infarcts while the sensitivity and specificity were estimated to be  66.8\% (95\% CI: 59.6-73.5)and 86.7\% (95\% CI: 81.8-90.7) respectively at the default threshold. The optimum operating threshold was determined using Youden’s index. At this optimum threshold, it was observed that the sensitivity of the DL-based algorithm improved to 80.2\% (95\% CI: 73.8 - 85.7) without substantial reduction in specificity 80.1\% (95\% CI: 74.5 - 84.9). For ICH, an AUC of 94.8\% (95\% CI: 87.4 - 100) was achieved. There was no change in sensitivity compared to the default and optimum threshold, while the specificity increased by 3\% using the optimum threshold. In contrast, the sensitivity of cranial fracture compared to the default and optimum threshold, an enhancement of 15.8\% was observed while the specificity decreased by 2.7\%. In Fig. \ref{auc_roc_plot}, the AUC-ROC plot for Cranial Fracture, ICH, and Infarct is given.
\begin{table}[!ht]
\centering
\caption{Performance Evaluation of DL-based algorithm at Default and Optimum Threshold; AUC: area under the curve; SN: sensitivity; SP: specificity, TP: true positive, TN: true negative, AP$^1$: all positive and AN$^1$: all negative}
\label{results_binary}
\begin{tabular}{@{}l|l|c|c|c|c@{}}
\midrule
\begin{tabular}[x]{@{}c@{}}\textbf{Target}\end{tabular} & \begin{tabular}[x]{@{}c@{}}\textbf{AUC (95\% CI) }\end{tabular} &\multicolumn{2}{ c |}{\textbf{Default threshold}}&\multicolumn{2}{ c }{\textbf{Optimum threshold}}\\
\cmidrule{3-6}
\textbf{Abnormality }&&\begin{tabular}[x]{@{}c@{}}SN (95\% CI) \\(TP/AP)\end{tabular}&
\begin{tabular}[x]{@{}c@{}}SP (95\% CI)\\(TN/AN)\end{tabular}&\begin{tabular}[x]{@{}c@{}}SN (95\% CI) \\(TP/AP)\end{tabular}&\begin{tabular}[x]{@{}c@{}}SP (95\% CI)\\(TN/AN)\end{tabular}\\
\midrule
Infarct & \begin{tabular}[x]{@{}c@{}}86.8\%\\(83.4-90.2)\end{tabular}& \begin{tabular}[x]{@{}c@{}}66.8\% \\(59.6-73.5) \\(125/187)\end{tabular} &
\begin{tabular}[x]{@{}c@{}}\textbf{86.7\%} \\(81.8-90.7) \\(209/241)\end{tabular}&
\begin{tabular}[x]{@{}c@{}}\textbf{80.2\%} \\(73.8-85.7) \\(150/187)\end{tabular}&
\begin{tabular}[x]{@{}c@{}}80.1\% \\(74.5-84.9)\\(193/241)\end{tabular}\\
\midrule
\begin{tabular}[x]{@{}c@{}}Intracranial \\hemorrhage\end{tabular}& \begin{tabular}[x]{@{}c@{}}\textbf{90.5\% }\\(69.6 – 98.8)\end{tabular}& \begin{tabular}[x]{@{}c@{}}90.5\%\\(69.6-98.8)\\(19/21)\end{tabular}&
\begin{tabular}[x]{@{}c@{}}93.1\%\\(90.2 – 95.4)\\(379/407)\end{tabular}&
\begin{tabular}[x]{@{}c@{}}\textbf{90.5\%}\\(69.6-98.8)\\(19/21)\end{tabular}&
\begin{tabular}[x]{@{}c@{}}\textbf{96.1\%}\\(93.7 – 97.7)\\(391/407)\end{tabular}\\
\midrule
\begin{tabular}[x]{@{}c@{}}Cranial\\ fracture\end{tabular}& \begin{tabular}[x]{@{}c@{}}95.2\%\\(90.1 – 100) \end{tabular}& 
\begin{tabular}[x]{@{}c@{}}78.3\%\\(56.3 – 92.5)\\(18/23)\end{tabular}&
\begin{tabular}[x]{@{}c@{}}\textbf{96.8\%}\\(94.6 – 98.3)\\(392/405)\end{tabular}&
\begin{tabular}[x]{@{}c@{}}\textbf{91.3\%}\\(72.0 – 98.9)\\(21/23)\end{tabular}&
\begin{tabular}[x]{@{}c@{}}94.1\%\\(91.3 – 96.2)\\(381/405)\end{tabular}\\
\midrule
\end{tabular}
\begin{tablenotes}\footnotesize
\item[$^1$] $AP = TP + FN $, $AP = TN + FP $.
\item[*] Default threshold: A threshold determined from internal testing results during model training.
\end{tablenotes}
\vspace{-5mm}
\end{table}
\subsection{Quantification of Infarct Volume}
The DL-based algorithm for identifying infarcts produces the infarct volume in mL. A total of 150 true positive scans for which both DL-based algorithms predicted volume and ground truth volume were available for this analysis. The reference standard was radiologist annotations done for each Head-NCCT scan images. 
\begin{table}[!ht]
\centering
\caption{Performance Evaluation of DL-based algorithm in Quantifying Infarct; MSE: Mean absolute error; STD: standard deviation.}
\label{results_quant}
\begin{tabular}{@{}c|l|l|l|c|c|c@{}}
\midrule
\textbf{Infarct Volume}&\textbf{NI}&\begin{tabular}[x]{@{}c@{}}\textbf{MAE}\end{tabular} &\begin{tabular}[x]{@{}c@{}}\textbf{Median}\end{tabular}&\begin{tabular}[x]{@{}c@{}}\textbf{STD}\end{tabular}&\multicolumn{2}{ c }{\textbf{Percentile}}\\
\cmidrule{6-7}
&&&&&\begin{tabular}[x]{@{}c@{}}10\% \end{tabular}&\begin{tabular}[x]{@{}c@{}}90\%\end{tabular}\\
\midrule
0 - 5 mL & 108&3.200 & 0.158 &1.121 &0.032&2.746\\
\midrule
$>$ 5 mL & 42&8.557 & 25.514 &34.751 &5.657&80.236\\
\midrule
Overall & 150 &4.700 & 0.600 &24.397&0.042&42.080\\
\midrule
\end{tabular}
\vspace{-5mm}
\end{table}
\begin{figure}[!htp]
\centering
\includegraphics[width=\textwidth]{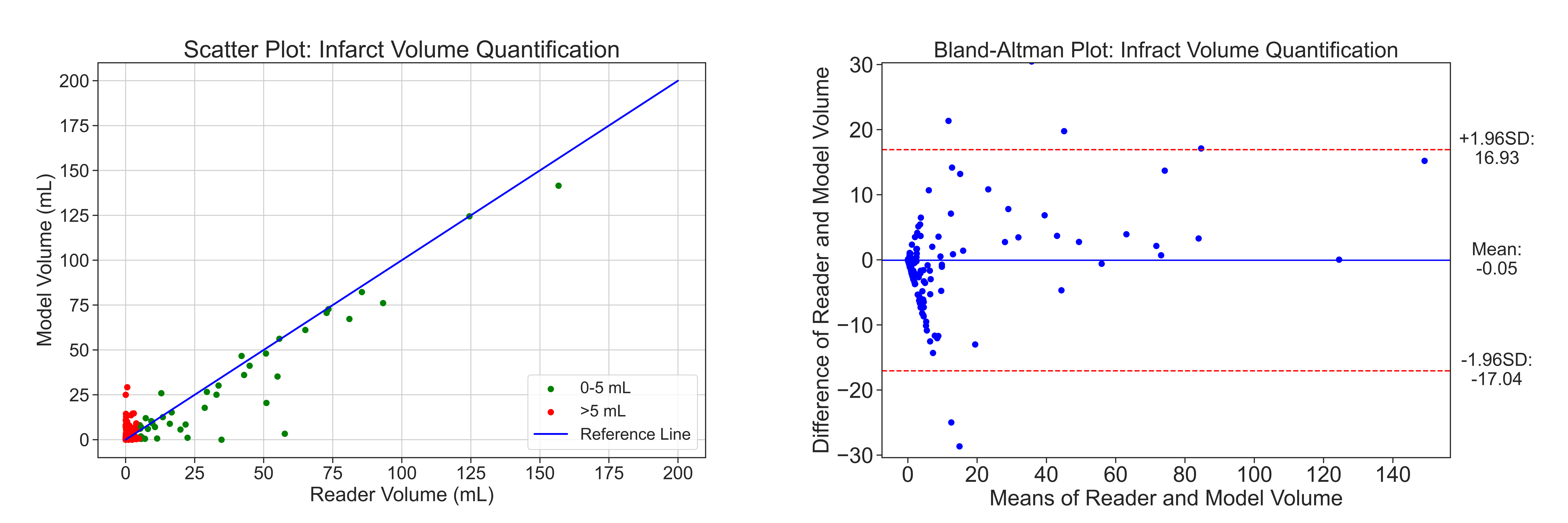}
\caption{1) Scatter plot for infarct volume quantification with volume distribution of 0-5 mL and $>$ 5 mL. 2) Bland-Altman plot for infarct volume quantification}
\label{quant_plot}
\end{figure}

The mean absolute error (MAE) was 4.7 mL for overall scans. Based on ground truth volume, the scans were further divided into two categories - scans with 0 - 5 mL and $>$ 5 mL infarcts volume, respectively. It can be observed from Table. \ref{results_quant} that the MAE for 0 - 5 mL and $>$ 5 mL scans were found to be 3.2 mL and 8.557 mL, respectively. In Fig. \ref{quant_plot} from the scatter plot of infarct volumes (1), it can be observed that with an increase in infarct volume, there is a positive correlation between DL-based algorithm volume and ground-truth annotated volume. The Bland-Altman plots showing good agreement between the ground truther annotation and predicted volume by the DL-based algorithm are shown in Fig. \ref{quant_plot} (2). 
\subsection{Visual Explanations of DL-based Algorithm}
\begin{figure}[!h]
\centering
\includegraphics[width=\textwidth]{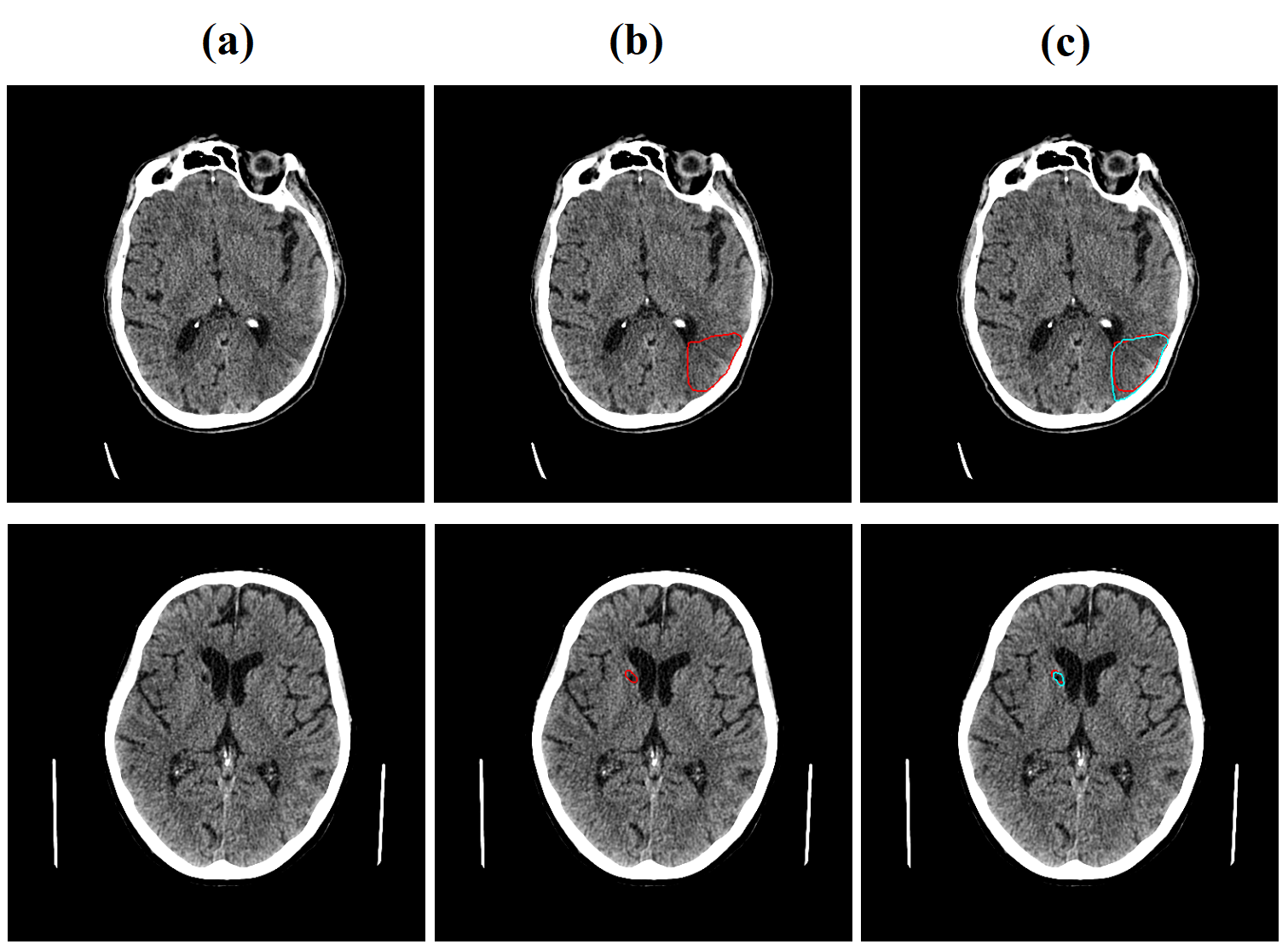}
\caption{Representative two head NCCT scan images with infarct. (a) The head NCCT scan with presence of infarct (b) The ground truther annotation around the region of infarct with red boundary (c) DL-based algorithm volume quantification (blue boundary) overlapping with ground truther annotation (red boundary).}
\label{repre_quant}
\end{figure}

The experimental findings depict that the evaluated DL-based algorithm achieved superior performance represented in Table. \ref{results_binary} and \ref{results_quant}. In most DL-based models the rationale behind the prediction is not reveled explicitly. Since these DL black box models can not be decomposed into intuitive and comprehensive modules, these models are hard to interpret. Consequently, the end-users develop skepticism and find the model difficult to trust. The emergence of explainable artificial intelligence (XAI) is an essential aspect of model transparency and the social right to explain DL inferences \cite{XAI_3},\cite{XAI_4}. XAI encompasses a better understanding of incoherent output, isolates failure modes, and builds trust in intelligent systems for effective incorporation into our everyday lives \cite{XAI_1}. The present evaluated DL-based algorithm outputs a boundary across the infarcts which revels rationale behind the superior performance.  In Fig. \ref{repre_quant} is can be observed that for both small and large infarcts volume on Head-NCCT scan, the model predicted boundary clearly overlaps with the ground truther boundary.

\section{Discussion}\label{sec12}
This retrospective study evaluated a deep learning algorithm for detecting infarcts in Head-NCCT scans. The algorithm had a good AUC of about 86\% in detecting infarcts. After adjusting for thresholds, a balanced sensitivity of 80.2\% and specificity of 80.1\% was estimated to detect infarcts. The algorithm's sensitivity in detecting infarcts in scans with no other target abnormalities was found to be 80\% (136 correctly detected out of 170). It did not differ from the overall sensitivity at optimum sensitivity. This states the robustness of the DL-based algorithm to identify infarcts with negligible drop in sensitivity with presence of other abnormalities. Additionally, it is to be noted that the sensitivity of Head-NCCT scans in detecting infarcts is generally considered low, especially in the case of hyperacute and acute ischemic strokes. In one study, the sensitivity of detecting acute ischemic stroke on head NCCT scans ranged from 57\% to 71\% with considerable inter-reader variability \cite{Dis_1}\cite{Dis_2}. Additionally, we evaluated the performance to detect ICH and cranial fracture, and both had excellent AUC. However, the interpretation is limited by low sample sizes for these two abnormalities. Our results also show that threshold adjustments might be needed before using such algorithms routinely for clinical decision support.

Deep learning or big data are often called "black box" and represent substantial obstacles in introducing intuitive and comprehensive modules into actual clinical practice; these models are challenging to interpret. However, the DL-based method validated in this study provides a post-hoc attention tool for the clinician to identify the lesion visually. In addition, the DL-based algorithm validated in this study encompasses a better understanding of incoherent output, isolates failure modes, and builds trust in intelligent systems for effective incorporation into routine clinical practice. Moreover, the proposed validation of the DL-based algorithm will be beneficial in the resource constraint areas with a limited number of radiologists or with only access to teleradiology facilities. 

Our study has limitations. First, the differentiation of infarct into acute and chronic infarct was not analyzed. Second, the ground truthing for the head NCCT scans images with the presence of infarcts was done by a single radiologist. Thirdly, there were not enough scans for the ICH and cranial fracture to estimate performance metrics with sufficient precision.
\section{Conclusion}\label{sec13}
The present study evaluated a DL-based algorithm to determine the presence and absence of ICH and infarcts on head-NCCT scans. The DL-based algorithm demonstrated high detection performance rate in identifying infarcts, ICH, and cranial fracture. Additionally, the DL-based algorithm exhibits a positive correlation between DL-based algorithm volume and ground-truth annotated volume. The study demonstrated the performance of ICH detection and infarcts detection and quantification to indicate the feasibility of introduction of such DL-algorithms in routine workflow in extensive healthcare facilities.

\section{Data Availability}
The datasets used or analyzed during the current study are available from the corresponding author on reasonable request.
\bibliography{sn-main}


\begin{thebibliography}{28}
\ifx \bisbn   \undefined \def \bisbn  #1{ISBN #1}\fi
\ifx \binits  \undefined \def \binits#1{#1}\fi
\ifx \bauthor  \undefined \def \bauthor#1{#1}\fi
\ifx \batitle  \undefined \def \batitle#1{#1}\fi
\ifx \bjtitle  \undefined \def \bjtitle#1{#1}\fi
\ifx \bvolume  \undefined \def \bvolume#1{\textbf{#1}}\fi
\ifx \byear  \undefined \def \byear#1{#1}\fi
\ifx \bissue  \undefined \def \bissue#1{#1}\fi
\ifx \bfpage  \undefined \def \bfpage#1{#1}\fi
\ifx \blpage  \undefined \def \blpage #1{#1}\fi
\ifx \burl  \undefined \def \burl#1{\textsf{#1}}\fi
\ifx \doiurl  \undefined \def \doiurl#1{\url{https://doi.org/#1}}\fi
\ifx \betal  \undefined \def \betal{\textit{et al.}}\fi
\ifx \binstitute  \undefined \def \binstitute#1{#1}\fi
\ifx \binstitutionaled  \undefined \def \binstitutionaled#1{#1}\fi
\ifx \bctitle  \undefined \def \bctitle#1{#1}\fi
\ifx \beditor  \undefined \def \beditor#1{#1}\fi
\ifx \bpublisher  \undefined \def \bpublisher#1{#1}\fi
\ifx \bbtitle  \undefined \def \bbtitle#1{#1}\fi
\ifx \bedition  \undefined \def \bedition#1{#1}\fi
\ifx \bseriesno  \undefined \def \bseriesno#1{#1}\fi
\ifx \blocation  \undefined \def \blocation#1{#1}\fi
\ifx \bsertitle  \undefined \def \bsertitle#1{#1}\fi
\ifx \bsnm \undefined \def \bsnm#1{#1}\fi
\ifx \bsuffix \undefined \def \bsuffix#1{#1}\fi
\ifx \bparticle \undefined \def \bparticle#1{#1}\fi
\ifx \barticle \undefined \def \barticle#1{#1}\fi
\bibcommenthead
\ifx \bconfdate \undefined \def \bconfdate #1{#1}\fi
\ifx \botherref \undefined \def \botherref #1{#1}\fi
\ifx \url \undefined \def \url#1{\textsf{#1}}\fi
\ifx \bchapter \undefined \def \bchapter#1{#1}\fi
\ifx \bbook \undefined \def \bbook#1{#1}\fi
\ifx \bcomment \undefined \def \bcomment#1{#1}\fi
\ifx \oauthor \undefined \def \oauthor#1{#1}\fi
\ifx \citeauthoryear \undefined \def \citeauthoryear#1{#1}\fi
\ifx \endbibitem  \undefined \def \endbibitem {}\fi
\ifx \bconflocation  \undefined \def \bconflocation#1{#1}\fi
\ifx \arxivurl  \undefined \def \arxivurl#1{\textsf{#1}}\fi
\csname PreBibitemsHook\endcsname

\bibitem[\protect\citeauthoryear{Fellows et~al.}{2005}]{Intro_1}
\begin{barticle}
\bauthor{\bsnm{Fellows}, \binits{L.K.}},
\bauthor{\bsnm{Heberlein}, \binits{A.S.}},
\bauthor{\bsnm{Morales}, \binits{D.A.}},
\bauthor{\bsnm{Shivde}, \binits{G.}},
\bauthor{\bsnm{Waller}, \binits{S.}},
\bauthor{\bsnm{Wu}, \binits{D.H.}}:
\batitle{Method matters: An empirical study of impact in cognitive
  neuroscience}.
\bjtitle{Journal of Cognitive Neuroscience}
\bvolume{17},
\bfpage{850}--\blpage{858}
(\byear{2005})
\end{barticle}
\endbibitem

\bibitem[\protect\citeauthoryear{Geva et~al.}{2011}]{Intro_2}
\begin{barticle}
\bauthor{\bsnm{Geva}, \binits{S.}},
\bauthor{\bsnm{Jones}, \binits{P.S.}},
\bauthor{\bsnm{Crinion}, \binits{J.T.}},
\bauthor{\bsnm{Price}, \binits{C.J.}},
\bauthor{\bsnm{Baron}, \binits{J.C.}},
\bauthor{\bsnm{Warburton}, \binits{E.A.}}:
\batitle{The neural correlates of inner speech defined by voxel-based
  lesion–symptom mapping}.
\bjtitle{Brain}
\bvolume{134},
\bfpage{3071}--\blpage{3082}
(\byear{2011})
\end{barticle}
\endbibitem

\bibitem[\protect\citeauthoryear{Utz et~al.}{2012}]{Intro_3}
\begin{barticle}
\bauthor{\bsnm{Utz}, \binits{S.}},
\bauthor{\bsnm{Humphreys}, \binits{G.W.}},
\bauthor{\bsnm{Chechlacz}, \binits{M.}}:
\batitle{Parietal substrates for dimensional effects in visual search: evidence
  from lesion-symptom mapping.}
\bjtitle{Brain : a journal of neurology}
\bvolume{136 Pt 3},
\bfpage{751}--\blpage{60}
(\byear{2012})
\end{barticle}
\endbibitem

\bibitem[\protect\citeauthoryear{Haji and Naval}{2020}]{Intro_5}
\begin{botherref}
\oauthor{\bsnm{Haji}, \binits{S.}},
\oauthor{\bsnm{Naval}, \binits{N.S.}}:
Management of intracerebral hemorrhage.
Evidence-Based Critical Care
(2020)
\end{botherref}
\endbibitem

\bibitem[\protect\citeauthoryear{Rymer}{2011}]{Intro_4}
\begin{barticle}
\bauthor{\bsnm{Rymer}, \binits{M.M.}}:
\batitle{Hemorrhagic stroke: intracerebral hemorrhage.}
\bjtitle{Missouri medicine}
\bvolume{108 1},
\bfpage{50}--\blpage{4}
(\byear{2011})
\end{barticle}
\endbibitem

\bibitem[\protect\citeauthoryear{Phipps and Cronin}{2020}]{Intro_6}
\begin{botherref}
\oauthor{\bsnm{Phipps}, \binits{M.S.}},
\oauthor{\bsnm{Cronin}, \binits{C.}}:
Management of acute ischemic stroke.
BMJ
\textbf{368}
(2020)
\end{botherref}
\endbibitem

\bibitem[\protect\citeauthoryear{Rorden et~al.}{2012}]{Intro_7}
\begin{barticle}
\bauthor{\bsnm{Rorden}, \binits{C.}},
\bauthor{\bsnm{Bonilha}, \binits{L.}},
\bauthor{\bsnm{Fridriksson}, \binits{J.}},
\bauthor{\bsnm{Bender}, \binits{B.}},
\bauthor{\bsnm{Karnath}, \binits{H.-O.}}:
\batitle{Age-specific ct and mri templates for spatial normalization}.
\bjtitle{NeuroImage}
\bvolume{61},
\bfpage{957}--\blpage{965}
(\byear{2012})
\end{barticle}
\endbibitem

\bibitem[\protect\citeauthoryear{Fiez et~al.}{2000}]{Intro_8}
\begin{botherref}
\oauthor{\bsnm{Fiez}, \binits{J.A.}},
\oauthor{\bsnm{Damasio}, \binits{H.}},
\oauthor{\bsnm{Grabowski}, \binits{T.J.}}:
Lesion segmentation and manual warping to a reference brain: Intra‐ and
  interobserver reliability.
Human Brain Mapping
\textbf{9}
(2000)
\end{botherref}
\endbibitem

\bibitem[\protect\citeauthoryear{Ashton et~al.}{2003}]{Intro_9}
\begin{botherref}
\oauthor{\bsnm{Ashton}, \binits{E.A.}},
\oauthor{\bsnm{Takahashi}, \binits{C.}},
\oauthor{\bsnm{Berg}, \binits{M.J.}},
\oauthor{\bsnm{Goodman}, \binits{A.}},
\oauthor{\bsnm{Totterman}, \binits{S.M.S.}},
\oauthor{\bsnm{Ekholm}, \binits{S.}}:
Accuracy and reproducibility of manual and semiautomated quantification of ms
  lesions by mri.
Journal of Magnetic Resonance Imaging
\textbf{17}
(2003)
\end{botherref}
\endbibitem

\bibitem[\protect\citeauthoryear{Pratiher and Chattoraj}{2018}]{Intro_10}
\begin{botherref}
\oauthor{\bsnm{Pratiher}, \binits{S.}},
\oauthor{\bsnm{Chattoraj}, \binits{S.}}:
Diving deep onto discriminative ensemble of histological hashing \&
  class-specific manifold learning for multi-class breast carcinoma taxonomy.
ICASSP 2019 - 2019 IEEE International Conference on Acoustics, Speech and
  Signal Processing (ICASSP),
1025--1029
(2018)
\end{botherref}
\endbibitem

\bibitem[\protect\citeauthoryear{Nawn et~al.}{2020}]{Intro_11}
\begin{barticle}
\bauthor{\bsnm{Nawn}, \binits{D.}},
\bauthor{\bsnm{Pratiher}, \binits{S.}},
\bauthor{\bsnm{Chattoraj}, \binits{S.}},
\bauthor{\bsnm{Chakraborty}, \binits{D.}},
\bauthor{\bsnm{Pal}, \binits{M.}},
\bauthor{\bsnm{Paul}, \binits{R.R.}},
\bauthor{\bsnm{Dutta}, \binits{S.}},
\bauthor{\bsnm{Chatterjee}, \binits{J.}}:
\batitle{Multifractal alterations in oral sub-epithelial connective tissue
  during progression of pre-cancer and cancer}.
\bjtitle{IEEE Journal of Biomedical and Health Informatics}
\bvolume{25},
\bfpage{152}--\blpage{162}
(\byear{2020})
\end{barticle}
\endbibitem

\bibitem[\protect\citeauthoryear{Govindarajan et~al.}{2022}]{Intro_12}
\begin{botherref}
\oauthor{\bsnm{Govindarajan}, \binits{A.}},
\oauthor{\bsnm{Govindarajan}, \binits{A.}},
\oauthor{\bsnm{Tanamala}, \binits{S.}},
\oauthor{\bsnm{Chattoraj}, \binits{S.}},
\oauthor{\bsnm{Reddy}, \binits{B.}},
\oauthor{\bsnm{Agrawal}, \binits{R.}},
\oauthor{\bsnm{Iyer}, \binits{D.}},
\oauthor{\bsnm{Srivastava}, \binits{A.}},
\oauthor{\bsnm{Kumar}, \binits{P.}},
\oauthor{\bsnm{Putha}, \binits{P.}}:
Role of an automated deep learning algorithm for reliable screening of
  abnormality in chest radiographs: A prospective multicenter quality
  improvement study.
Diagnostics
\textbf{12}
(2022)
\end{botherref}
\endbibitem

\bibitem[\protect\citeauthoryear{Pratiher et~al.}{2018}]{Intro_13}
\begin{botherref}
\oauthor{\bsnm{Pratiher}, \binits{S.}},
\oauthor{\bsnm{Chattoraj}, \binits{S.}},
\oauthor{\bsnm{Mukherjee}, \binits{R.}}:
Stationplot: A new non-stationarity quantification tool for detection of
  epileptic seizures.
2018 IEEE Global Conference on Signal and Information Processing (GlobalSIP),
499--503
(2018)
\end{botherref}
\endbibitem

\bibitem[\protect\citeauthoryear{Chilamkurthy et~al.}{2018}]{Intro_14}
\begin{barticle}
\bauthor{\bsnm{Chilamkurthy}, \binits{S.}},
\bauthor{\bsnm{Ghosh}, \binits{R.}},
\bauthor{\bsnm{Tanamala}, \binits{S.}},
\bauthor{\bsnm{Biviji}, \binits{M.}},
\bauthor{\bsnm{Campeau}, \binits{N.G.}},
\bauthor{\bsnm{Venugopal}, \binits{V.K.}},
\bauthor{\bsnm{Mahajan}, \binits{V.}},
\bauthor{\bsnm{Rao}, \binits{P.}},
\bauthor{\bsnm{Warier}, \binits{P.}}:
\batitle{Deep learning algorithms for detection of critical findings in head ct
  scans: a retrospective study}.
\bjtitle{The Lancet}
\bvolume{392},
\bfpage{2388}--\blpage{2396}
(\byear{2018})
\end{barticle}
\endbibitem

\bibitem[\protect\citeauthoryear{Wilke et~al.}{2011}]{Intro_15}
\begin{barticle}
\bauthor{\bsnm{Wilke}, \binits{M.}},
\bauthor{\bsnm{Haan}, \binits{B.}},
\bauthor{\bsnm{Juenger}, \binits{H.}},
\bauthor{\bsnm{Karnath}, \binits{H.-O.}}:
\batitle{Manual, semi-automated, and automated delineation of chronic brain
  lesions: A comparison of methods}.
\bjtitle{NeuroImage}
\bvolume{56},
\bfpage{2038}--\blpage{2046}
(\byear{2011})
\end{barticle}
\endbibitem

\bibitem[\protect\citeauthoryear{Rekik et~al.}{2012}]{Intro_16}
\begin{barticle}
\bauthor{\bsnm{Rekik}, \binits{I.}},
\bauthor{\bsnm{Allassonni{\`e}re}, \binits{S.}},
\bauthor{\bsnm{Carpenter}, \binits{T.}},
\bauthor{\bsnm{Wardlaw}, \binits{J.M.}}:
\batitle{Medical image analysis methods in mr/ct-imaged acute-subacute ischemic
  stroke lesion: Segmentation, prediction and insights into dynamic evolution
  simulation models. a critical appraisal}.
\bjtitle{NeuroImage : Clinical}
\bvolume{1},
\bfpage{164}--\blpage{178}
(\byear{2012})
\end{barticle}
\endbibitem

\bibitem[\protect\citeauthoryear{Chan}{2007}]{Intro_17}
\begin{barticle}
\bauthor{\bsnm{Chan}, \binits{T.}}:
\batitle{Computer aided detection of small acute intracranial hemorrhage on
  computer tomography of brain}.
\bjtitle{Computerized medical imaging and graphics : the official journal of
  the Computerized Medical Imaging Society}
\bvolume{31 4-5},
\bfpage{285}--\blpage{98}
(\byear{2007})
\end{barticle}
\endbibitem

\bibitem[\protect\citeauthoryear{Liu et~al.}{2008}]{Intro_18}
\begin{botherref}
\oauthor{\bsnm{Liu}, \binits{R.}},
\oauthor{\bsnm{Tan}, \binits{C.L.}},
\oauthor{\bsnm{Leong}, \binits{T.-Y.}},
\oauthor{\bsnm{Lee}, \binits{C.K.}},
\oauthor{\bsnm{Pang}, \binits{B.C.}},
\oauthor{\bsnm{Lim}, \binits{C.C.T.}},
\oauthor{\bsnm{Tian}, \binits{Q.}},
\oauthor{\bsnm{Tang}, \binits{S.}},
\oauthor{\bsnm{Zhang}, \binits{Z.}}:
Hemorrhage slices detection in brain ct images.
2008 19th International Conference on Pattern Recognition,
1--4
(2008)
\end{botherref}
\endbibitem

\bibitem[\protect\citeauthoryear{Matesin et~al.}{2001}]{Intro_19}
\begin{botherref}
\oauthor{\bsnm{Matesin}, \binits{M.}},
\oauthor{\bsnm{Lonvaric}, \binits{S.}},
\oauthor{\bsnm{Petravic}, \binits{D.}}:
A rule-based approach to stroke lesion analysis from ct brain images.
ISPA 2001. Proceedings of the 2nd International Symposium on Image and Signal
  Processing and Analysis. In conjunction with 23rd International Conference on
  Information Technology Interfaces (IEEE Cat.,
219--223
(2001)
\end{botherref}
\endbibitem

\bibitem[\protect\citeauthoryear{Obuchowski}{1994}]{stats_1}
\begin{barticle}
\bauthor{\bsnm{Obuchowski}, \binits{N.A.}}:
\batitle{Computing sample size for receiver operating characteristic studies}.
\bjtitle{Investigative Radiology}
\bvolume{29},
\bfpage{238}--\blpage{243}
(\byear{1994})
\end{barticle}
\endbibitem

\bibitem[\protect\citeauthoryear{Zhou et~al.}{2002}]{stats_2}
\begin{bchapter}
\bauthor{\bsnm{Zhou}, \binits{X.-H.}},
\bauthor{\bsnm{Obuchowski}, \binits{N.A.}},
\bauthor{\bsnm{McClish}, \binits{D.K.}}:
\bctitle{Statistical methods in diagnostic medicine}.
(\byear{2002})
\end{bchapter}
\endbibitem

\bibitem[\protect\citeauthoryear{Clopper and Pearson}{1934}]{Stats_3}
\begin{barticle}
\bauthor{\bsnm{Clopper}, \binits{C.J.}},
\bauthor{\bsnm{Pearson}, \binits{E.S.}}:
\batitle{The use of confidence or fiducial limits illustrated in the case of
  the binomial}.
\bjtitle{Biometrika}
\bvolume{26},
\bfpage{404}--\blpage{413}
(\byear{1934})
\end{barticle}
\endbibitem

\bibitem[\protect\citeauthoryear{DeLong et~al.}{1988}]{Stats_4}
\begin{barticle}
\bauthor{\bsnm{DeLong}, \binits{E.R.}},
\bauthor{\bsnm{DeLong}, \binits{D.M.}},
\bauthor{\bsnm{Clarke‐Pearson}, \binits{D.L.}}:
\batitle{Comparing the areas under two or more correlated receiver operating
  characteristic curves: a nonparametric approach.}
\bjtitle{Biometrics}
\bvolume{44 3},
\bfpage{837}--\blpage{45}
(\byear{1988})
\end{barticle}
\endbibitem

\bibitem[\protect\citeauthoryear{Do{\v{s}}ilovi{\'c} et~al.}{2018}]{XAI_3}
\begin{bchapter}
\bauthor{\bsnm{Do{\v{s}}ilovi{\'c}}, \binits{F.K.}},
\bauthor{\bsnm{Br{\v{c}}i{\'c}}, \binits{M.}},
\bauthor{\bsnm{Hlupi{\'c}}, \binits{N.}}:
\bctitle{Explainable artificial intelligence: A survey}.
In: \bbtitle{2018 41st International Convention on Information and
  Communication Technology, Electronics and Microelectronics (MIPRO)},
pp. \bfpage{0210}--\blpage{0215}
(\byear{2018}).
\bcomment{IEEE}
\end{bchapter}
\endbibitem

\bibitem[\protect\citeauthoryear{Adadi and Berrada}{2018}]{XAI_4}
\begin{barticle}
\bauthor{\bsnm{Adadi}, \binits{A.}},
\bauthor{\bsnm{Berrada}, \binits{M.}}:
\batitle{Peeking inside the black-box: A survey on explainable artificial
  intelligence (xai)}.
\bjtitle{IEEE Access}
\bvolume{6},
\bfpage{52138}--\blpage{52160}
(\byear{2018})
\end{barticle}
\endbibitem

\bibitem[\protect\citeauthoryear{Gunning}{2017}]{XAI_1}
\begin{botherref}
\oauthor{\bsnm{Gunning}, \binits{D.}}:
Explainable artificial intelligence (xai).
Defense Advanced Research Projects Agency (DARPA), nd Web
\textbf{2}(2)
(2017)
\end{botherref}
\endbibitem

\bibitem[\protect\citeauthoryear{Broderick et~al.}{2007}]{Dis_1}
\begin{barticle}
\bauthor{\bsnm{Broderick}, \binits{J.P.}},
\bauthor{\bsnm{Connolly}, \binits{S.E.}},
\bauthor{\bsnm{Feldmann}, \binits{E.}},
\bauthor{\bsnm{Hanley}, \binits{D.F.}},
\bauthor{\bsnm{Kase}, \binits{C.S.}},
\bauthor{\bsnm{Krieger}, \binits{D.}},
\bauthor{\bsnm{Mayberg}, \binits{M.R.}},
\bauthor{\bsnm{Morgenstern}, \binits{L.B.}},
\bauthor{\bsnm{Ogilvy}, \binits{C.S.}},
\bauthor{\bsnm{Vespa}, \binits{P.M.}},
\bauthor{\bsnm{Zuccarello}, \binits{M.}}:
\batitle{Guidelines for the management of spontaneous intracerebral hemorrhage
  in adults. 2007 update. guideline from the american heart
  association/american stroke association stroke council, high blood pressure
  research council, and the quality of care and outcomes in research
  interdisciplinary working gro}.
\bjtitle{Stroke}
\bvolume{38},
\bfpage{2001}--\blpage{2023}
(\byear{2007})
\end{barticle}
\endbibitem

\bibitem[\protect\citeauthoryear{Srinivasan et~al.}{2006}]{Dis_2}
\begin{barticle}
\bauthor{\bsnm{Srinivasan}, \binits{A.}},
\bauthor{\bsnm{Goyal}, \binits{M.}},
\bauthor{\bsnm{Azri}, \binits{F.A.}},
\bauthor{\bsnm{Lum}, \binits{C.}}:
\batitle{State-of-the-art imaging of acute stroke.}
\bjtitle{Radiographics : a review publication of the Radiological Society of
  North America, Inc}
\bvolume{26 Suppl 1},
\bfpage{75}--\blpage{95}
(\byear{2006})
\end{barticle}
\endbibitem

\end{thebibliography}

\end{document}